\documentclass[preprint,preprintnumbers, prd, floatfix, superscriptaddress,nofootinbib] {revtex4-1}
\usepackage{epsfig}
\usepackage{subfigure}
\usepackage{dcolumn}
\usepackage{bm}
\usepackage[usenames ,dvipsnames]{xcolor}
\usepackage{slashed}
\usepackage{graphicx,color}
\usepackage[bookmarksnumbered, pdfstartview=FitH,colorlinks,urlcolor=blue, citecolor=blue,linkcolor=blue] {hyperref}
\begin{document}
\title{Investigating the $\eta^\prime_{1}(1855)$ exotic state in the $J/\psi\to\eta^\prime_{1}(1855)\eta^{(\prime)}$ decays}

\author{Yao Yu}
\email{Corresponding author: yuyao@cqupt.edu.cn}
\affiliation{Chongqing University of Posts \& Telecommunications, Chongqing, 400065, China}
 \affiliation{Department of Physics and Chongqing Key Laboratory for Strongly Coupled Physics, Chongqing University, Chongqing 401331, People's Republic of China}
\author{Zhuang Xiong}
\affiliation{Chongqing University of Posts \& Telecommunications, Chongqing, 400065, China}
\author{Han Zhang}
\affiliation{School of Physics and Microelectronics,
Zhengzhou University, Zhengzhou, Henan 450001, China}
\author{Bai-Cian Ke}
\email{Corresponding author: baiciank@ihep.ac.cn}
\affiliation{School of Physics and Microelectronics,
Zhengzhou University, Zhengzhou, Henan 450001, China}
\author{Yi Teng}
\affiliation{Chongqing University of Posts \& Telecommunications, Chongqing, 400065, China}
\author{Qing-Shan Liu}
\affiliation{Chongqing University of Posts \& Telecommunications, Chongqing, 400065, China}
\author{Jia-Wei Zhang}
\email{Corresponding author: jwzhang@cqust.edu.cn}
\affiliation{Department of Physics, Chongqing University of Science and Technology, Chongqing, 401331,
 China}
\date{\today}

\begin{abstract}
  An analysis of the $J/\psi\to \eta\eta^\prime\gamma$ decay by the BESIII
  collaboration claims the observation of an exotic state $\eta_1(1855)$ with
  $I^GJ^{PC}=0^+1^{-+}$. To establish its C-parity partner
  $\eta^\prime_{1}(1855)$ in the picture of the $K \bar{K}_1(1400)$ molecular
  state, we propose that $J/\psi\to\eta^\prime_{1}(1855)\eta^{(\prime)}$
  receives the main contributions from the final state interaction of
  $KK^*$($K^+ K^{*-}$, $K^- K^{*+}$, $K^0 \bar{K}^{*0}$, and $\bar{K}^0 K^{*0}$).
  Specifically, $K$ and $K^*$ in $J/\psi\to KK^*$ decays transform as
  $\eta^\prime_{1}(1855)\eta^{(\prime)}$, by exchanging a $K_1(1400)$. We
  predict
  ${\cal B}(J/\psi\to \eta^{\prime}_1(1855)\eta)=(6.3^{+12.6}_{-3.5})\times 10^{-6}$,
  and
  ${\cal B}(J/\psi\to \eta^{\prime}_1(1855)\eta^{\prime}) =(6.5^{+6.6}_{-4.6})\times 10^{-6}$,
  which can be studied in the $J/\psi\to K^*\bar{K}^*\eta^{(\prime)}$ decays.
\end{abstract}
\maketitle
\section{introduction}

Although most conventional hadrons are mesons or baryon,
Quantum Chromodynamics actually allows the existence of other types of states,
called exotic states as long as the color confinement is satisfied. One
decisive way to judge whether a meson is exotic states or not is to examine its
$J^{PC}$, for which conventional mesons can't have quantum numbers $J^{PC}=0^{--}$,
$(\textrm{even})^{+-}$, and $(\textrm{odd})^{-+}$. The BESIII Collaboration has
recently observed a new state $\eta_{1}(1855)\equiv \eta_{1}$ with quantum
numbers $J^{PC} = 1^{-+}$ on the $\eta\eta^\prime$ invariant mass spectrum of
the $J/\psi\to \eta\eta^\prime\gamma$ decay~\cite{BESIII:2022riz,BESIII:2022riz2}
and determined the mass and width to be
\begin{eqnarray}\label{MandW}
m_{\eta_1}&=&(1.855\pm 0.009^{+0.006}_{-0.001})~\text{GeV}\,,\,\Gamma_{\eta_1}=(188\pm 18^{+3}_{-8})~\text{MeV}\,.
\end{eqnarray}
The $J^{PC}$ of $\eta_{1}$ unambiguously indicates it is an exotic state.
However, it deserves more efforts to further determine which type of exotic
states the $\eta_{1}$ is.

Many theoretical hypotheses interpreting the nature of $\eta_{1}$ have been
proposed immediately after its observation, such as an $\bar{s}sg$ isoscalar
hybrid meson~\cite{Shastry:2023ths, Qiu:2022ktc,Chen:2022qpd,Shastry:2022mhk,Chen:2022isv} or a
tetraquark state~\cite{Wan:2022xkx}, but the mass' being around the threshold
of total mass of $K$ and $\bar{K}_1(1400)\equiv \bar{K}_1$ makes the $\eta_{1}$
more naturally to be interpreted as a $K \bar{K_1}$+c.c.~molecular
state~\cite{Dong:2022cuw,Yang:2022lwq,Wang:2022sib}.
($KK_1$ denotes the various combinations $K^+K_1^{-}$, $K^-K_1^{+}$,
$K^0\bar{K}_1^{0}$, and $\bar{K}^0 K_1^{0}$ in the following.)
Reference~\cite{Dong:2022cuw} has showed the binding energies of the
isoscalar $K\bar{K}_1(1400)$ are all negative in various situations in its
Fig.~2 and proved the attractive force between $K$ and $\bar{K}_1$, by
exchanging mesons, is strong enough to form a bound state using the
one-boson-exchange model. This shows the newly discovered $\eta_{1}$ could be
the candidate of a $KK_1$ molecular state with $J^{PC} = 1^{-+}$. At the same
time, the molecular model uniquely predicted that $\eta_{1}(1855)$ should have
a C-parity partner with $J^{PC}=1^{--}$, called
$\eta^{\prime}_{1}(1855)\equiv \eta^{\prime}_{1}$~\cite{Dong:2022cuw, Wang:2022sib}.
Hence, examining existence of the $\eta^{\prime}_{1}$ is very important to
decide whether the $\eta^{(\prime)}_{1}$ are molecular states or not.

In this paper, we analyze the productions of $\eta^{(\prime)}_{1}$, assuming
they are $KK_1$ bound states, in the $J/\Psi\to \eta^\prime_{1}\eta^{(\prime)}$
decays. In principle, all the possible bases that can connect the initial state
$J/\Psi$ and final state $\eta^\prime_{1}\eta^{(\prime)}$ should be considered.
The direct estimations of these production process at the quark level are
difficult, but the loops composed by hadrons can be regarded as the major
contributions as indicated in Refs.~\cite{Wu:2021caw,Wu:2021cyc}. The dominant
diagrams contributing to $J/\Psi\to \eta^\prime_{1}\eta^{(\prime)}$, as
depicted in Fig.~\ref{triangle}, can give large enough branching ratios to be
observed. The $\eta^\prime_{1}$ could be produced through the final state
interaction in the $J/\psi\to K K^*$ decay ($K K^*= K^+ K^{*-}$, $K^- K^{*+}$,
$K^0 \bar{K}^{*0}$, and $\bar{K}^0 K^{*0}$), followed by the $K$-$K^*$
rescattering. The $K$ and $K^*$ then transform to
$\eta^\prime_{1}\eta^{(\prime)}$ with the $K_1$ exchange in the
triangle-rescattering process. The size of the contribution from this
triangle-rescattering effect highly depends on the couplings of involved
intermediate interactions, which are crucial and required in calculation of
the triangle loop. Fortunately, the branching fractions of the
$J/\psi\to KK^*$ and $K_1\to K^*\pi$ decays have been measured to be at
$10^{-2}$ level and almost 100\%,
respectively~\cite{ParticleDataGroup:2020ssz}, implying a strong
coupling constant $g_{K_1 K^* \eta^{(\prime)}}$ with the helping of the SU(3)
flavour symmetry. In addition, the $\eta^\prime_1$, as a candidate of a
$KK_1$ molecular, should couple to $KK_1$ strongly~\cite{Dong:2022cuw}.
Therefore, we investigate the $J/\psi\to \eta_{1}^{\prime} \eta^{(\prime)}$
decays in the molecular model in this work and show that they are anticipated
to be accessible in the BESIII experiment.

\section{Formalism}
\begin{figure}[t!]
\includegraphics[width=3.2in]{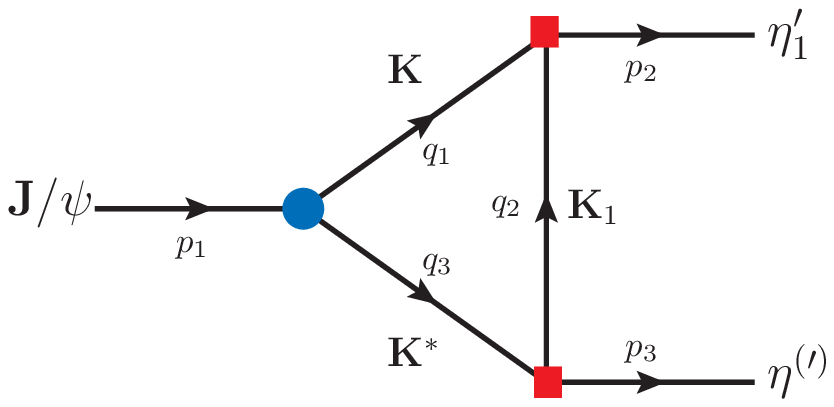}
\caption{Rescattering $J/\psi\to\eta^\prime_{1}\eta^{(\prime)}$ decays.}\label{triangle}
\end{figure}

In this section, we analyze the $J/\psi\to \eta^\prime_{1}\eta^{(\prime)}$
decay in the molecular model. Its triangle-rescattering process, as depicted in
Fig.~\ref{triangle}, can be separated into three parts:$J/\psi\to KK^*$,
$\eta_1^\prime\to  KK_1$, and $K_1\to  K^{*}\eta^{(\prime)}$.

The first part is $J/\psi\to KK^*$. The relevant Lagrangian term
is~\cite{Haber:1985cv,Huang:2013wpa}
  \begin{eqnarray}
  {\cal L}_1 &=& \frac{g_{\psi KK^*}}{m_{\psi}}K\epsilon_{\mu\nu\rho\sigma}\partial^{\mu}\psi^{\nu}\partial^{\rho}K^{*\sigma}\,,
\end{eqnarray}
where $g_{\psi K K^*}$ is the coupling constant for the $J/\psi \to KK^*$
decay and $\epsilon^{\nu}_{\psi}(\epsilon^{\nu}_{K^*})$ is the polarization
four-vector of the $J/\psi(K^*)$. We derive this amplitude to be
\begin{eqnarray}\label{weakamp1}
{\cal M}_1(J/\psi\to K K^*)
&=&\frac{g_{\psi KK^*}}{m_{\psi}}\epsilon_{\mu\nu\rho\sigma}p^{\mu}_{\psi}\epsilon^{\nu}_{\psi}p^{\rho}_{K^*}\epsilon^{*\sigma}_{K^*}\,.
\end{eqnarray}
The relations between various
$g_{\psi KK^*}$ can be given by SU(3) flavour symmetry~\cite{Haber:1985cv,Huang:2013wpa}:
\begin{eqnarray}\label{weakamp2}
g_{\psi K^+K^{*-}}&=&g_{\psi K^-K^{*+}}=g_{8}-\frac{g^{M}_{88}}{2\sqrt{3}}+\frac{|g^{E}_{88}|}{3}e^{i\delta_E},\nonumber\\
g_{\psi K^0\bar{K}^{*0}}&=&g_{\psi \bar{K}^0 K^{*0}}=g_{8}-\frac{g^{M}_{88}}{2\sqrt{3}}-\frac{2|g^{E}_{88}|}{3}e^{i\delta_E}\,,
\end{eqnarray}
where $g_{8}$, $g^{M}_{88}$, $g^{E}_{88}$ and $\delta_E$ are the coupling
constants of the octet term, the mass-breaking term, and the
electromagnetic-breaking term, and the phase angle between electromagnetic and
strong interaction, respectively.

The second part is $\eta_1^\prime\to  KK_1$. The corresponding Lagrangian term
is
  \begin{eqnarray}
   {\cal L}_2&=& g_{\eta^\prime_1KK_1}\Phi[-p_E^2]K\eta^\prime_1\cdot K_1\,,
\end{eqnarray}
and the amplitude is derived to be
\begin{eqnarray}\label{strong}
{\cal M}_2(\eta^\prime_1\to  KK_1)&=&g_{\eta^\prime_1KK_1}\Phi[-p_E^2]\epsilon_{\eta^\prime_1}\cdot\epsilon_{K_1}\,,
\end{eqnarray}
where $p_E$ is the Euclidean Jacobi momentum and
$g_{\eta^\prime_1KK_1}= g_{\eta^\prime_1K^+K_1^{*-}}=g_{\eta^\prime_1K^-K_1^{*+}}=g_{\eta^\prime_1K^0\bar{K}_1^{*0}}=g_{\eta^\prime_1\bar{K}^0K_1^{*0}}$
are the coupling constants of the $\eta^\prime_1\to KK_1$ decays.
For the sake of calculation, $p_E^2=-(p_{K_1}-\lambda p_{\eta^\prime_1})^2$
and $\lambda=\frac{m_{K}}{m_{K_1}+m_K}$. The correlation function
$\Phi[-p_E^2]$ can be parameterized as a Gaussian form vertex
function~\cite{Branz:2007xp,Chen:2015igx} or a pole form vertex
function~\cite{Giacosa:2007bn,Giacosa:2012de,Schneitzer:2014rsa}:
\begin{eqnarray}
  \Phi[-p_E^2] &=& \exp[-p_E^2/\Lambda_{1}^2]\,,\,\,\,\,\textrm{[Gaussian form]}\nonumber\\
  \Phi[-p_E^2] &=& 1+p_E^2/\Lambda_{1}^2\,,\,\,\,\,\,\,\,\,\,\,\,\textrm{[Pole form]}
\end{eqnarray}
with $\Lambda_{1}$ is the size parameter. In this paper, the Gaussian form will
cause the integration divergence and, alternatively, we choose the pole vertex
form correlation function. We determine the coupling constants between the
hadronic molecule and its components using the pole vertex form by the
consequence of the $K\ddot{a}$ll$\acute{e}$n-Lehmann representation (see
Eqs.~\ref{a1}-\ref{a6}) and the triangle diagram can be calculate by the
$'$tHooft-Veltman technique~\cite{tHooft:1978jhc} (see Eq.~\ref{int2}).

Note that using the Gaussian form is more common than the Pole form to
determine the coupling constant by the compositeness condition in an one loop
diagram~\cite{Weinberg:1962hj}. However, calculating triangle diagram with the
Gaussian form may have singularities, called anomalous
thresholds~\cite{Ivanov:2003ge}. A parameter $z_{loc}$ has been
proposed~\cite{Branz:2009cd} to prejudge whether there are anomalous
thresholds. The parameter $z_{loc}$ describes the triangle diagram in which $A$
particle decay into $B$ and $C$ particles with $a$, $b$, and $c$ particles as
propagators. Take Fig.~\ref{triangle} for example,
$A/B/C=J/\Psi, \eta_1^{\prime}, \eta^{(\prime)}$ and $a/b/c=K/K_1/K^*$. Then,
$z_{loc}$ is given by
\begin{eqnarray}
    z_{loc} &=& \alpha_am_a^2+ \alpha_bm_b^2+ \alpha_cm_c^2-\alpha_a\alpha_cM_A^2-\alpha_a\alpha_bM_B^2-\alpha_b\alpha_cM_C^2,
\end{eqnarray}
with $\alpha_a+\alpha_b+\alpha_c=1$ and $\alpha_i\geq 0\,(i=a,b,c)$.
There will be no anomalous thresholds if $z_{loc}$ is always positive.\footnote{
This could happen when, for example, $A$ is a molecular state and the circumstance
$|m_a-m_c|<m_A<m_a+m_c$ is satisfied.}
In this paper, the set of masses
$(M_A,M_B,M_C,m_a,m_b.m_c)=(m_{J/\Psi},m_{\eta_1^{\prime}},m_{\eta^{(\prime)}},m_K,m_{K_1},m_{K^*})$
make $z_{loc}$ not always positive, implying divergence will happen if the
Gaussian form is used.

The third part is $K_1\to  K^{*}\eta^{(\prime)}$, whose Lagrangian term and
amplitude are written as
  \begin{eqnarray}
  {\cal L}_3 &=& g_{K_1K^{*}\eta^{(\prime)}}\eta^{(\prime)}K_1\cdot K^*\,,
\end{eqnarray}
and
\begin{eqnarray}
  {\cal M}_3(K_1\to  K^{*}\eta^{(\prime)}) &=&-i g_{K_1K^{*}\eta^{(\prime)}}\epsilon_{K_1}\cdot\epsilon_{K^*}\,,
\end{eqnarray}
where
\begin{eqnarray}
  g_{K^{-}_1K^{*-}\eta} &=& \frac{b}{4}\cos\theta\cos\phi-\frac{a}{4}\sin\theta\cos\phi+ \frac{b}{2\sqrt{2}}\cos\theta\sin\phi+ \frac{a}{2\sqrt{2}}\sin\theta\sin\phi \nonumber\\
   g_{K^{-}_1K^{*-}\eta^{\prime}} &=& \frac{b}{2\sqrt{2}}\cos\theta\cos\phi+\frac{a}{2\sqrt{2}}\sin\theta\cos\phi- \frac{b}{4}\cos\theta\sin\phi+ \frac{a}{4}\sin\theta\sin\phi
\end{eqnarray}
and
\begin{eqnarray}
  g_{K^{-}_1K^{*-}\eta^{(\prime)}} &=& g_{K^{+}_1K^{*+}\eta^{(\prime)}}=g_{K^{0}_1K^{*0}\eta^{(\prime)}}=g_{\bar{K}^{0}_1\bar{K}^{*0}\eta^{(\prime)}}\,,
\end{eqnarray}
where ($a$, $b$) are the parameters for coupling constants and ($\theta$,
$\phi$) are the mixing angles of ($K_1\sim K_1^\prime$,
$\eta\sim \eta^\prime$)~\cite{Divotgey:2013jba}.

Eventually, the amplitude of the triangle-rescattering process for the
$J/\psi\to \eta^\prime_{1}\eta^{(\prime)}$ decay is obtained by
\begin{eqnarray}\label{w1}
{\cal M}(J/\psi\to \eta_1^{\prime}\eta^{(\prime)})&=&\sum^{4}_{i=1}\int \frac{d^4{q}_{1}}{(2\pi)^{4}}
\frac{{\cal M}_1 {\cal M}_2 {\cal M}_3 F_{\Lambda_{K_1}}(q_{2}^2)}
{(q_{1}^{2}-m_{K}^{2})(q_{2}^{2}-m_{K_1}^{2})(q_{3}^{2}-m_{K^{*}}^{2})}\,,
\end{eqnarray}
where $\sum^{4}_{i=1}$ sums over all possible Feynman diagrams for
$K^{+(*)}_{(1)}$, $K^{-(*)}_{(1)}$, $K^{0(*)}_{(1)}$, and $\bar{K}^{0(*)}_{(1)}$,
and
$F_{\Lambda_{K_1}}(q_{2}^2)\equiv(\Lambda_2^{2}-m^{2}_{K_1})/(\Lambda^{2}_2-q^{2}_{2})$
is the monopole form factor~\cite{Tornqvist:1993ng,Li:1996yn,Yu:2020vlt},
which can be adopted to represent the off-shell effect by exchanging $K_1$
mesons and also plays the role of avoiding integration divergences. Besides,
$q_2=p_{\eta^\prime_1}-q_1$ and $q_3=p_\psi-q_1$ correspond to the momentum
flows in Fig.~\ref{triangle}. In the general form, one can expresses the
amplitude as
\begin{eqnarray}\label{w2}
{\cal M}(J/\psi\to \eta_1^{\prime}\eta^{(\prime)})&=&-i\frac{g_{\psi\eta_1^{\prime}\eta^{(\prime)}}}{m_{\psi}}\epsilon_{\mu\nu\rho\sigma}p^{\mu}_{\psi}\epsilon^{\nu}_{\psi}p^{\rho}_{\eta_1^{\prime}}\epsilon^{*\sigma}_{\eta_1^{\prime}}\,.
\end{eqnarray}
To obtain $g_{\psi\eta_1^{\prime}\eta^{(\prime)}}$, one needs to integrate
over the variables of the triangle loop in Eq.~\ref{w1}, which gives
\begin{eqnarray}\label{w3}
  &{\cal M}&(J/\psi\to \eta_1^{\prime}\eta^{(\prime)}) =\nonumber\\
  && -i\frac{g_{\eta_1KK_{1}}}{4\pi^2m_{\psi}}\left(g_8-\frac{g^{88}_M}{2\sqrt{3}}-\frac{g^{88}_E}{6}e^{i\delta_E}\right)\Lambda^2_{1}
g_{K_1K^{*}\eta^{(\prime)}}\epsilon_{\mu\nu\rho\sigma}p^{\mu}_{\psi}\epsilon^{\nu}_{\psi}(D^{\rho}-D^{'\rho})\epsilon^{*\sigma}_{\eta_1^{\prime}}\,.
\end{eqnarray}
The propagators of $K^*$ and $K_1$ are supposed to contain vector and tensor
structures, but contributions from the tensor structures will be zero due to
the term $\epsilon_{\mu\nu\rho\sigma}$ in Eq.~\ref{w1}. Hence, we only
consider the vector structures here. The vector four-point function is
written as
\begin{eqnarray}
 D^{\rho}&=&
\int \frac{d^{4}q_1}{i \pi^2}\frac{q_1^{\rho}}
{(q_1^{2}-m_{K}^{2}+i\epsilon)[q_2^{2}-m_{K_1}^{2}+i\epsilon][q_3^{2}-m_{K^*}^{2}+i\epsilon][(q_1-\lambda p_{\eta_1^{\prime}})^{2}-\Lambda_{1}^{2}+i\epsilon]}\,,
\end{eqnarray}
such that one obtains
\begin{eqnarray}\label{cij_1}
D^{\rho}&=&p_{\eta_1^{\prime}}^\rho D_1+ p_{\psi}^\rho D_2\,,
\end{eqnarray}
with the linear combination of scalar point functions
\begin{eqnarray}\label{int1}
D_1&\equiv&\frac{m^{2}_{\psi}+m^{2}_{\eta^{(\prime)}}-m^2_{\eta_1^{\prime}}}{m^{2}_{\psi}m^2_{\eta_1^{\prime}}-(m^{2}_{\psi}+m^{2}_{\eta^{(\prime)}}-m^2_{\eta_1^{\prime}})^2}[(m^{2}_{\psi}+m^{2}_{K}-m^2_{K_1})D_0+C_{0A}-C_{0B}]\,.
\end{eqnarray}
The $D_2$ term doesn't contributed to
${\cal M}(J/\psi\to \eta_1^{\prime}\eta^{(\prime)})$ due to
$\epsilon_{\mu\nu\alpha\beta}p_{\psi}^\mu p_{\psi}^\nu=0$. The one loop
scalar 3- and 4-point functions can be calculated with the
$'$tHooft-Veltman
technique~\cite{tHooft:1978jhc,Hahn:1998yk,Denner:2005nn,Hsiao:2019ait,Hsiao:2021tyq,Yu:2021euw}
and are given by
\begin{eqnarray}\label{int2}
D_{0}&=&
\int \frac{d^{4}q_1}{i \pi^2}\frac{1}
{(q_1^{2}-m_{K}^{2}+i\epsilon)[q_2^{2}-m_{K_1}^{2}+i\epsilon][q_3^{2}-m_{K^*}^{2}+i\epsilon][(q_1-\lambda p_{\eta_1^{\prime}})^{2}-\Lambda_{1}^{2}+i\epsilon]}\,,\nonumber\\
C_{0A}&=&
\int \frac{d^{4}q_1}{i \pi^2}\frac{1}
{[q_2^{2}-m_{K_1}^{2}+i\epsilon][q_3^{2}-m_{K^*}^{2}+i\epsilon][(q_1-\lambda p_{\eta_1^{\prime}})^{2}-\Lambda_{1}^{2}+i\epsilon]}\,,\nonumber\\
C_{0B}&=&
\int \frac{d^{4}q_1}{i \pi^2}\frac{1}
{(q_1^{2}-m_{K}^{2}+i\epsilon)[q_3^{2}-m_{K^*}^{2}+i\epsilon][(q_1-\lambda p_{\eta_1^{\prime}})^{2}-\Lambda_{1}^{2}+i\epsilon]}\,.
\end{eqnarray}
As for the term $D^{\prime \rho}$, one can obtain
$D^{\prime \rho}=p_{\eta_1^{\prime}}^\rho D_1^\prime+ p_{\psi}^\rho D_2^\prime$
by replacing $m_{K_1}$ in Eqs.~(\ref{cij_1}-\ref{int2}) with $\Lambda_2$. Next,
one can derive $g_{\psi\eta_1^{\prime}\eta^{(\prime)}}$ by comparing
Eqs.~\ref{w2} and~\ref{w3} and defining $\tilde{D}_{1}=D_{1}-D'_{1}$:
\begin{eqnarray}\label{cij_2}
g_{\psi\eta_1^{\prime}\eta^{(\prime)}}&=& -\frac{g_{\eta_1KK_{1}}}{4\pi^2}\left(g_8-\frac{g^{88}_M}{2\sqrt{3}}-\frac{g^{88}_E}{6}e^{i\delta_E}\right)\Lambda^2_{1}
g_{K_1K^{*}\eta^{(\prime)}}\tilde{D}_{1}\,.
\end{eqnarray}
At this point, all parameters relevant to
${\cal M}(J/\psi\to \eta_1^{\prime}\eta^{(\prime)})$ are given except
$g_{\eta^{\prime}_1KK_{1}}$, which can be determined by a consequence of the
$K\ddot{a}$ll$\acute{e}$n-Lehmann representation (see the discussion below).

\begin{figure}[t!]
\includegraphics[width=3.2in]{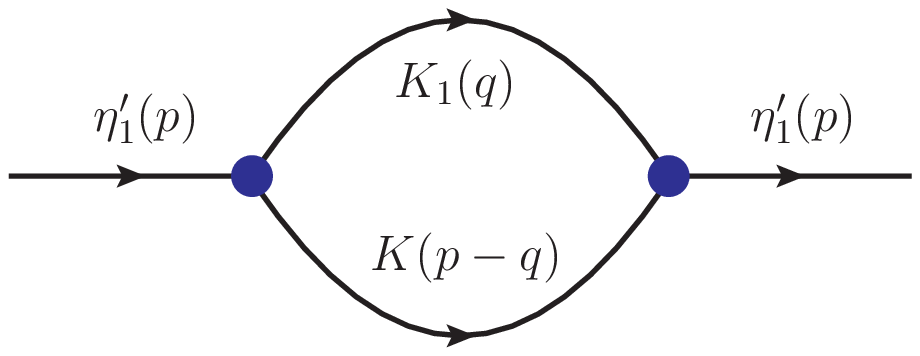}
\caption{the one-loop correction  to the propagator of $\eta^\prime _{1} $.}\label{oneloop}
\end{figure}
Upon resummation of the one-loop contributions, as in Fig.~\ref{oneloop}, the
propagator of $\eta^\prime_1$ takes the form
\begin{eqnarray}
  D^{\mu\nu}_{\eta^\prime_1} &=&-i \frac{g^{\mu\nu}+...}{p^2-m^2_{\eta^\prime_1}+Re[\Sigma(m^2_{\eta^\prime_1})]- \Sigma(p^2)+i\epsilon}\,.
\end{eqnarray}
The metric tensor term, $g^{\mu\nu}$, already provides enough information to
determine the coupling $g_{\eta^{\prime}_1KK_{1}}$ and the rest can be ignored
(denoted as [...]). The spectral function of the state $\eta_1^\prime$ in the
$K\ddot{a}$ll$\acute{e}$n-Lehmann representation can be obtained as the
imaginary part of the propagator~\cite{Giacosa:2007bn,Giacosa:2012de}:
\begin{eqnarray}\label{a1}
  d_{\eta^\prime_1}(p^2) &=& \frac{1}{\pi}|\lim_{\epsilon\rightarrow 0^+} Im{[p^2-m^2_{\eta^\prime_1}+Re[\Sigma(m^2_{\eta^\prime_1})]- \Sigma(p^2)+i\epsilon]^{-1}}|
\end{eqnarray}
and the normalization is required to be satisfied:
\begin{eqnarray}\label{a2}
  \int^{+\infty}_{0}d_{\eta^\prime_1}(p^2)d p^2&=&1\,.
\end{eqnarray}
In the above equation, we define
\begin{eqnarray}\label{a3}
 \Sigma^{\mu\nu}(p^2) &=& -g^{\mu\nu}\Sigma(p^2)+...\,,\nonumber\\
 \Sigma^{\mu\nu}(p^2) &=& i\int \frac{d^{4}q}{(2\pi)^4}\frac{(-g^{\mu\nu}+q^\mu q^\nu/m_{K_1}^{2})\Phi^2[(q-(1-\lambda) p)^2]}
{(q^{2}-m_{K_1}^{2}+i\epsilon)[(q- p)^{2}-m_{K}^{2}+i\epsilon]}\,,
 \end{eqnarray}
and can give
\begin{eqnarray}\label{a4}
  \Sigma(p^2) &=&-\frac{2}{3}I^\prime_3-\frac{(p^2+m_{K_1}^{2}-m_{K}^{2})^2}{12p^2m_{K_1}^{2}}I^\prime_3-\frac{I^\prime_{2a}}{m_{K_1}^{2}}\left(\frac{3-2\lambda}{12}+\frac{m_{K_1}^{2}-m_{K}^{2}}{12p^2}+\frac{m_{K_1}^{2}-\Lambda_{1}^{2}}{6(1-\lambda) p^2}\right)\nonumber\\
  &+&\frac{I^\prime_{2b}}{m_{K_1}^{2}}\left(\frac{1-2\lambda}{12}+\frac{m_{K_1}^{2}-m_{K}^{2}}{12p^2}+\frac{\Lambda_{1}^{2}-m_{K}^{2}}{6\lambda p^2}\right)+\frac{I_{2a}-I_{2c}}{6(1-\lambda)m_{K_1}^{2} p^2}+\frac{I_{2b}-I_{2c}}{6\lambda m_{K_1}^{2} p^2}\,,
\end{eqnarray}
where
\begin{eqnarray}\label{a5}
  I_3 &=&\frac{g^2_{\eta^\prime_1KK_{1}}}{\pi^2}\int \frac{d^{4}q}{i \pi^2}\frac{1}
{(q^{2}-m_{K_1}^{2}+i\epsilon)\{[q-(1-\lambda)p]^{2}-\Lambda_{1}^{2}+i\epsilon\}[(q- p)^{2}-m_{K}^{2}+i\epsilon]}\,,\nonumber\\
   I_{2a} &=&\frac{g^2_{\eta^\prime_1KK_{1}}}{\pi^2}\int \frac{d^{4}q}{i \pi^2}\frac{1}
{(q^{2}-m_{K_1}^{2}+i\epsilon)\{[q-(1-\lambda)p]^{2}-\Lambda_{1}^{2}+i\epsilon\}}\,,\nonumber\\
I_{2b} &=&\frac{g^2_{\eta^\prime_1KK_{1}}}{\pi^2}\int \frac{d^{4}q}{i \pi^2}\frac{1}
{\{[q-(1-\lambda)p_{\eta^\prime_1}]^{2}-\Lambda_{1}^{2}+i\epsilon\}[(q- p)^{2}-m_{K}^{2}+i\epsilon]}\,,\nonumber\\
I_{2c} &=&\frac{g^2_{\eta^\prime_1KK_{1}}}{\pi^2}\int \frac{d^{4}q}{i \pi^2}\frac{1}
{(q^{2}-m_{K_1}^{2}+i\epsilon)[(q- p)^{2}-m_{K}^{2}+i\epsilon]}\,,
\end{eqnarray}
and
\begin{eqnarray}\label{a6}
I^\prime_3 =\frac{dI_3}{d\Lambda_{1}^{2}}\,,\,\,\,I^\prime_{2a(b)}=\frac{dI_{2a(b)}}{d\Lambda_{1}^{2}}\,.
\end{eqnarray}
The normalization of the spectral function of the $\eta_1^\prime$,
Eq.~\ref{a2}, causes a constraint in the $K\ddot{a}$ll$\acute{e}$n-Lehmann
representation. With this constraint, the $g_{\eta^{\prime}_1KK_{1}}$ term in
Eq.~\ref{a5} can be evaluated analytically with the $'$tHooft-Veltman
technique.
More specifically, one can obtain an analytical expression for
$d_{\eta^\prime_1}(p^2)$ by substituting Eqs.~\ref{a3}-\ref{a6} into Eq.~\ref{a1}.
After integrating out the momentum $p^2$ in Eq.~\ref{a2}, one can determine the
relationship between $g_{\eta^{\prime}_1KK_{1}}$ and $\Lambda_{1}$, as shown in
Fig.~\ref{spec1}. In this relationship, the masses of the relevant mesons are the
only input parameters.

Note that $I_{2a}$, $I_{2b}$, or $I_{2c}$ itself possesses logarithmic
divergence. However, Eq.~\ref{a4} doesn't diverge because the divergence will
cancel by $I_{2a}-I_{2c}$ and $I_{2b}-I_{2c}$. 

\section{Numerical results and Discussions}
In the numerical analysis, we adopt
$(g_{8}$, $g^{M}_{88}$, $g^{E}_{88})$=$(6.57\pm0.16,1.81\pm0.41,1.73\pm0.22)\times10^{-3}$,
$\delta_E=(69.5\pm12.6)^\circ$ from Ref.~\cite{Huang:2013wpa}\footnote{Parameters
$(g_{8}$, $g^{M}_{88}$, $g^{E}_{88})$, and $\delta_E$ in SU(3) flavor symmetry theory
describe the $J/\psi$ to pseudoscalar-vector decays, e.g.~$J/\psi\to\rho\pi$,
$J/\psi\to\omega\pi$, and $J/\psi\to K^*K$. They are extracted from fitting to
measured branching ratios.},
$(a,b)=(5.43\pm0.26,7.00\pm0.22)$~GeV, and $\phi=(56.3\pm4.2)^\circ$ from
Ref.~\cite{Divotgey:2013jba}. 

\begin{figure}[t!]
  \includegraphics[width=2.5in]{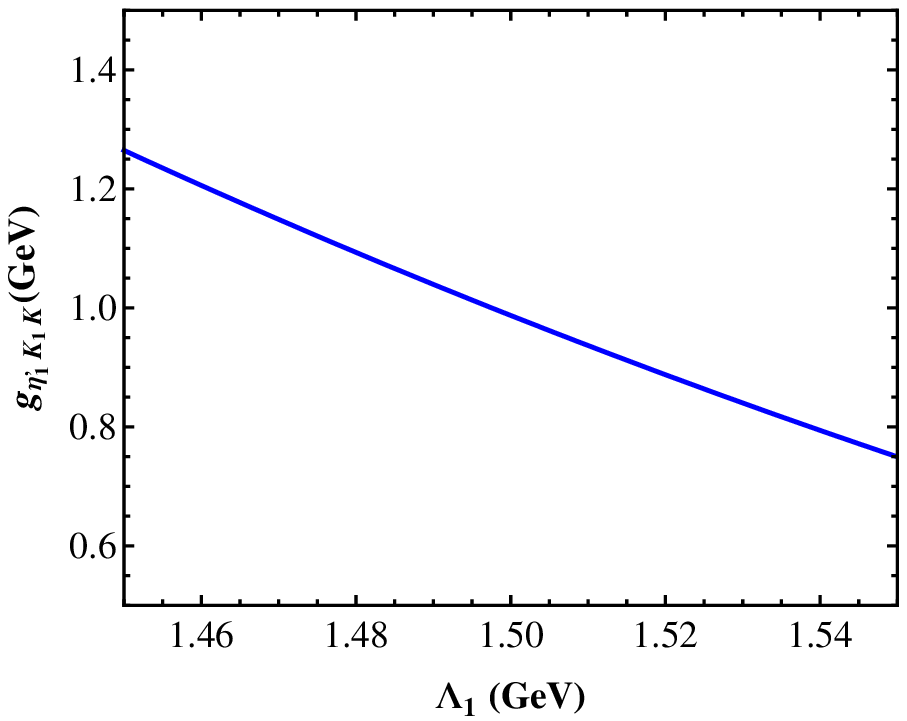}
  \caption{The coupling constant $g_{\eta^\prime_1KK_{1}}$ related to the
    parameter $\Lambda_{1}$.}\label{spec1}
\end{figure}

We can then fit $g_{\eta^{\prime}_1K_1K}=1.00^{+0.26}_{-0.25}$~GeV using
$\Lambda_1=1.50^{+0.05}_{-0.05}$~GeV~\cite{Giacosa:2007bn,Giacosa:2012de, Schneitzer:2014rsa}
and Eqs.~\ref{a1}-\ref{a6}. The choice of $\Lambda_1$ are typical in hadronic
theories. Figure~\ref{spec1} shows the coupling constant
$g_{\eta^\prime_1KK_{1}}$ related to $\Lambda_{1}$. Empirically, we are
allowed to use $\Lambda_2\simeq2\sim3$~GeV, which is not sensitive for the
branching ratio of $J/\psi\to \eta^{\prime}_1\eta^{(\prime)}$ because the
triangle Feynman diagrams to calculate
${\cal M}(J/\psi\to \eta^\prime_1\eta^{(\prime)})$ is convergent even without
the monopole form factor. As a consequence, we obtain
\begin{eqnarray}\label{pre_br}
{\cal B}(J/\psi\to \eta^{\prime}_1\eta)
&=&(6.3^{+12.6}_{-3.5})\times 10^{-6}\,,\,{\cal B}(J/\psi\to \eta^{\prime}_1\eta^{\prime})
=(6.5^{+6.6}_{-4.6})\times 10^{-6}\,.
\end{eqnarray}
Approximately, we present the resonant branching fractions as
\begin{eqnarray}
{\cal B}(J/\psi\to \eta^{\prime}_1\eta^{(\prime)},\eta^{\prime}_1\to K^*\bar{K}^*)
&\simeq&
{\cal B}(J/\psi\to \eta^{\prime}_1\eta^{(\prime)}){\cal B}(\eta^{\prime}_1\to K^*\bar{K}^*)\,,
\end{eqnarray}
and, with ${\cal B}(\eta^{\prime}_1\to K^*\bar{K}^*)\simeq22\%$~\cite{Dong:2022cuw},
predict
\begin{eqnarray}
{\cal B}(J/\psi\to  \eta^{\prime}_1\eta,\eta^{\prime}_1\to K^*\bar{K}^*)
&=&(1.4^{+2.8}_{-0.8})\times 10^{-6}\,,\nonumber\\
{\cal B}(J/\psi\to  \eta^{\prime}_1\eta^{\prime},\eta^{\prime}_1\to K^*\bar{K}^*)
&=&(1.4^{+1.5}_{-1.0})\times 10^{-6}\,.
\end{eqnarray}
The uncertainty is evaluated by repeating the calculation after varying the
parameters according to their uncertainties. The uncertainty associated with
$\Lambda_1$ dominates. The larger $\Lambda_1$ yields the lower predicted
branching fractions, but when $\Lambda_1$ is set to be in the typical region,
the predicted branching fractions remains being of $\mathcal{O}(10^{-6})$.


\section{Conclusions}
The C-parity partner of $\eta_{1}$, called $\eta^{\prime}_1$, is a peculiar
prophecy in the molecular model. Confirming the existence of the
$\eta^{\prime}_1$ will be a critical support of the molecular model and helps
to pin down the nature of $\eta_{1}$.
On the other side, the absence of $\eta^{\prime}_1$ would correspond to a
falsification of the molecular approach.
We have studied the rescattering decays
$J/\psi\to K K^*\to\eta^{\prime}_1\eta^{(\prime)}$. In the triangle loop, $K$
and $K^*$ transform as $\eta^\prime_{1}$ and $\eta^{(\prime)}$ by exchanging
$K_1$, respectively. We have proposed the
$J/\psi\to \eta^{\prime}_1\eta^{(\prime)},\,\eta^{\prime}_1\to K^*\bar{K}^*$
decay as a candidate decay to search for the $\eta^{\prime}_1$ exotic state.
In particular, we have predicted
${\cal B}(J/\psi\to \eta^{\prime}_1\eta)=(6.3^{+12.6}_{-3.5})\times 10^{-6}$ and
${\cal B}(J/\psi\to \eta^{\prime}_1\eta^{\prime})=(6.5^{+6.6}_{-4.6})\times 10^{-6}$
in the molecular model.

The BESIII collaboration has collected about 10 billion $J/\psi$ events at
$\sqrt{s}=3.097$~GeV and can reach the sensitivity of $10^{-7}$-$10^{-6}$
level for branching fractions of $J/\psi$ decays. Our proposal is shown to be
accessible by the BESIII experiment.

\section*{ACKNOWLEDGMENTS}
We would like to thank Dr.~Zhi Yang, Xu-Chang Zheng, and Rui-Yu Zhou for useful discussions.
YY was supported in part by National Natural Science Foundation of China (NSFC) under Contracts No.~11905023, No.~12047564 and No.~12147102, the Natural Science Foundation of Chongqing (CQCSTC) under Contracts No.~cstc2020jcyj-msxmX0555 and the Science and Technology Research Program of Chongqing Municipal Education Commission under Contracts No.~KJQN202200605 and No.~KJQN202200621; BCK was supported in part by NSFC under Contracts No.~11875054 and No.~12192263 and Joint Large-Scale Scientific Facility Fund of the NSFC and the Chinese Academy of Sciences under Contracts No.~U2032104; JWZ was supported in part by NSFC under Contracts No.~12275036, CQCSTC under Contracts No.~cstc2021jcyj-msxmX0681 and the Science and Technology Research Program of Chongqing Municipal Education Commission under Contracts No.~KJQN202001541.

\end{document}